# Oxidized organic molecules in the tropical free troposphere over Amazonia

Qiaozhi Zha[1], Diego Aliaga[1], Radovan Krejci[2], Victoria Sinclair[1], Cheng Wu[2], Wiebke Scholz[3], Liine Heikkinen[1,2], Eva Partoll[3], Yvette Gramlich[2], Wei Huang[1], Markus Leiminger[3,4], Joonas Enroth[1], Otso Peräkylä[1], Runlong Cai[1], Xuemeng Chen[1], Alkuin Maximilian Koenig[5], Fernando Velarde[5], Isabel Moreno[5], Tuukka Petäjä[1], Paulo Artaxo[6], Paolo Laj[1,7], Armin Hansel[3], Samara Carbone[8], Markku Kulmala[1,9,10], Marcos Andrade[5,11], Douglas Worsnop[1,12], Claudia Mohr[2], Federico Bianchi[1,*]

**Affiliations**

[1] Institute for Atmospheric and Earth System Research /Physics, University of Helsinki, Helsinki, Finland

[2] Department of Environmental Science & Bolin Centre for Climate Research, Stockholm University, Stockholm, Sweden

[3] Institute for Ion and Applied Physics, University of Innsbruck, Innsbruck, Austria

[4] Ionicon Analytik Ges.m.b.H., Innsbruck, Austria

[5] Laboratory for Atmospheric Physics, Institute for Physics Research, Universidad Mayor de San Andrés, La Paz, Bolivia

[6] Institute of Physics, University of Sao Paulo, Sao Paulo, SP, Brazil

[7] University of Grenoble Alpes, CNRS, IRD, Grenoble-INP, IGE (UMR 5001), Grenoble, France

[8] Federal University of Uberlândia, Uberlândia, MG, Brazil

[9] Joint International Research Laboratory of Atmospheric and Earth System Sciences, Nanjing University, Nanjing, China

[10] Beijing Advanced Innovation Center for Soft Matter Science and Engineering, Beijing University of Chemical Technology, Beijing, China

[11] Department of Atmospheric and Oceanic Sciences, University of Maryland, College Park, MD, USA

[12] Aerodyne Research, Inc., Billerica, MA, USA

*Corresponding author. Email: federico.bianchi@helsinki.fi

**Abstract**

New particle formation (NPF) in the tropical free troposphere (FT) is a globally important source of cloud condensation nuclei, affecting cloud properties and climate. Oxidized organic molecules (OOMs) produced from biogenic volatile organic compounds are believed to contribute to aerosol formation in the tropical FT, but without direct chemical observations. We performed in-situ molecular-level OOMs measurements at the Bolivian station Chacaltaya at 5240 meters above sea level, on the western edge of Amazonia. For the first time, we demonstrate the presence of OOMs, mainly with 4-5 carbon atoms, simultaneously in both gas and particulate phases in tropical FT air from Amazonia. These observations, combined with air mass history analyses, indicate that the observed OOMs are linked to isoprene emitted from the rainforests hundreds of kilometers away. Based on particle-phase measurements, we find that these compounds can contribute to the growth of newly formed particles, and are potentially crucial for new particle formation in the tropical free troposphere on a continental scale. Our study will thus improve the understanding of aerosol formation process in the tropics.

**Introduction**

The tropical free troposphere (FT) can host large numbers of aerosol particles, serving as cloud condensation nuclei (CCN), and thus affects the climate system on a global scale (Williamson et al., 2019; Weigel et al., 2011). Atmospheric new particle formation (NPF) has been consistently observed in the tropical FT in regions with high concentrations (in terms of particle number) of ultrafine particles (UFPs, here defined as particles with diameters between 10-50 nm) using aircraft and is thus likely to be a significant source of FT aerosols (Williamson et al., 2019; Weigel et al., 2011; Wang et al., 2016; Andreae et al., 2018). Oxidized organic molecules (OOMs), produced from biogenic volatile organic compounds (BVOCs) carried up to the FT by mesoscale convective systems, are hypothesized to be key components forming aerosols due to their reduced volatility at low temperatures (Williamson et al., 2019; Weigel et al., 2011; Andreae et al., 2018; Kulmala et al., 2006; Palmer et al., 2022).

Recent modeling studies have reported that biogenic-related NPF likely dominates FT aerosol formation in tropical BVOC emission hotspots like Amazonia (Zhao et al., 2020; Palmer et al., 2022). Elucidating the chemical composition of OOMs is required to constrain model simulations and improve understanding of the mechanism and influence of biogenic-related NPF in the tropical FT (Zhao et al., 2020). Direct observations to date are limited to airborne studies lacking necessary chemistry

instrumentation(Williamson et al., 2019; Weigel et al., 2011; Wang et al., 2016; Andreae et al., 2018). The Southern hemisphere high ALTitude Experiment on particle Nucleation And growth (SALTENA) campaign (see (Bianchi et al., 2021) and Methods) performed direct molecular-level observations of OOMs using a set of state-of-the-art mass spectrometers at the Bolivian Global Atmosphere Watch (GAW) station Chacaltaya (CHC; 5240 m above sea level), on the western edge of the Amazon Basin (Fig. S1). These measurements complemented long-term observations at CHC of, e.g., particle number size distribution and equivalent black carbon (eBC). For this campaign, we reconstructed the 96-hour air mass history using the Lagrangian particle dispersion model FLEXPART-WRF to determine the origin and footprint of air masses arriving at CHC (see (Aliaga et al., 2021) and Methods).

**Results**

The present study focuses on measurements in January 2018 during the austral summer (wet season), when FT air from Amazonia exerted a large influence on CHC compared to other periods of the campaign (Methods). The data were divided into FT, mixed FT and local air (non-FT), and daytime events. The FT events occurred during nighttime (19:00–06:00; local time, UTC - 4) when the influence of FT air was dominant at CHC. These periods were characterized by low water vapor mixing ratio (WVMR) and eBC concentrations, as tropical FT air is typically drier and less polluted (here indicated by the lower eBC concentration) than the boundary layer (BL) air during the wet season (see Methods for the detailed description of FT event identification) (Wiedensohler et al., 2018; Chauvigne et al., 2019; Sun and Lindzen, 1993). The remaining nighttime periods were defined as non-FT periods when the influence of local BL air was more significant compared to FT periods. Due to the thermally-driven local mixing layer cycle and air circulation, the impact of BL air was evident at CHC from 07:00 to 18:00. As a result, during daytime periods, CHC was impacted by urban air pollution from the nearby La Paz – El Alto metropolitan area (Wiedensohler et al., 2018; Chauvigne et al., 2019).

During the study period, FT events at CHC (Table S1) were frequently observed (on 14 out of 17 nights) under the persistent influence of air masses from Amazonia. An FT event is exemplarily shown in Fig. 1 for the night of 10 January 2018 (from 21:10 to 00:30), when very low WVMR (4.7 g kg$^{-1}$) and eBC concentrations (close to the detection limit) were measured. Detailed, high-resolution analysis of the modeled source-receptor relationship (SRR; see (Aliaga et al., 2021) and Methods) for this event, which links the sampled air masses at CHC via their transport and residence time therein, shows the Amazon Basin as the origin (Fig. 1a and S2, with hourly air mass history from 19:00 to 03:00). The SRR vertical profile (Fig. 1b) further shows that a large fraction of these air masses had spent considerable time within the Amazon BL in the region of 800 - 1400 km away from CHC before ascending to the FT. Similar to all other FT events in January 2018, MODIS (Moderate Resolution Image Spectroradiometer) satellite images show that several mesoscale convective systems were present concurrently over this region (Fig. S3 and S4), indicating air masses from the Amazon BL were transported to the tropical FT through convective lifting. Once in the FT, the air masses were transported via horizontal advection to CHC in ≤36 hours. In this way, the Amazon Basin can continuously and widely affect the tropical FT during the wet season.

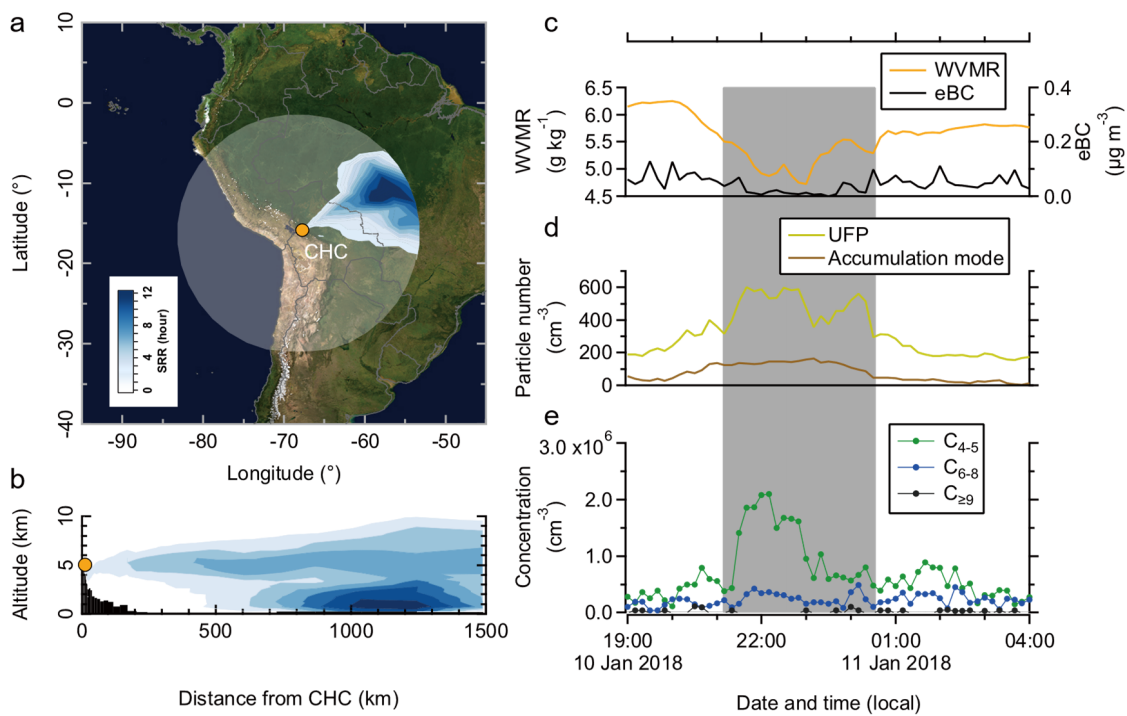

**Figure 1. An Amazon FT event observed at CHC in the night of 10 January 2018. a**, Horizontal profile of the vertically integrated source-receptor relationship (SRR, units of hours) derived from the FLEXPART-WRF model and averaged from 21:00 to 00:00 (local date and time, UTC-4). The semi-transparent circle indicates the horizontal output domain of the model. The model output is in 1-hour time resolution. The color bar denotes the SRR values of the passive air tracers. **b**, Vertical profile of the SRR integrated over the radial direction, averaged from 21:00 to 00:00. Black shading indicates the topography near the station. **c**, Concentrations of WVMR and eBC. The grey shaded area denotes the exact period of the FT event, identified with WVMR ≤5.5 g kg$^{-1}$ and eBC ≤0.08 μg m$^{-3}$ (Methods). **d**, Concentrations of UFPs (10 - 50 nm diameter) and accumulation mode particles (100 - 500 nm diameter). **e**, Concentrations of oxidized organic molecules (OOMs) measured by a nitrate-based CI-APi-TOF; grouped based on their number of carbon atoms ($C_{4-5}$, $C_{6-8}$, and $C_{\geq 9}$). $C_{\geq 9}$ OOM concentration is below

the detection limit of the CI-APi-TOF (Methods) for most of the time during this FT event.

The composition of FT air from Amazonia showed notable differences from that during non-FT periods on the night of 10 January 2018. WVMR and eBC concentrations started to decrease at 20:00 and reached a minimum after about one hour, indicating that by 21:10, the CHC was dominantly influenced by FT air (Fig. 1c). The FT condition persisted for 3 hours until shortly after midnight on 11 January 2018. Fig. 1d shows increased UFP concentrations during that period. The hourly averaged particle size spectra (Fig. S5) show that the increase in UFPs was primarily due to enhancement of particles with diameters between 20 – 30 nm. This is consistent with particle size spectra measured in the Amazon FT using aircraft (Wang et al., 2016; Andreae et al., 2018). We divide observed OOMs (Fig. 1e) into three groups based on their number of carbon atoms ($C_{4-5}$, $C_{6-8}$, and $C_{\geq 9}$). Whereas $C_{6-8}$ and $C_{\geq 9}$ OOM concentrations remained at low levels throughout the night, $C_{4-5}$ OOM concentrations increased during the FT event. A concurrent increase in the total signal of organic ions was also measured by the atmospheric pressure interface time-of-flight mass spectrometer (APi-TOF; Fig. S6), as compositions of OOMs and organic ions (negative ion adducts of OOMs) were usually identical (Bianchi et al., 2016, 2017). Two more FT events on the night of 11 and 18 January are shown in Fig. S7 and S8.

Overall, the characteristics determined based on all FT events (Table S1) are distinct from those of non-FT periods. WVMR and eBC concentrations (Fig. 2a, b) were significantly lower during FT conditions compared to non-FT conditions. High number concentrations of UFPs (Fig. 2c, d) are another typical feature of tropical FT air (Wang et al., 2016; Andreae et al., 2018; Williamson et al., 2019). Concentrations of $C_{6-8}$ and $C_{\geq 9}$ OOMs were much lower than $C_{4-5}$ OOM concentrations (Fig. 2e, f, g), and showed no clear differences between FT events and non-FT periods for these two groups. In contrast, $C_{4-5}$ OOM concentrations in FT events were significantly higher than that during non-FT periods, indicating tropical FT air is enriched in $C_{4-5}$ OOMs during the wet season.

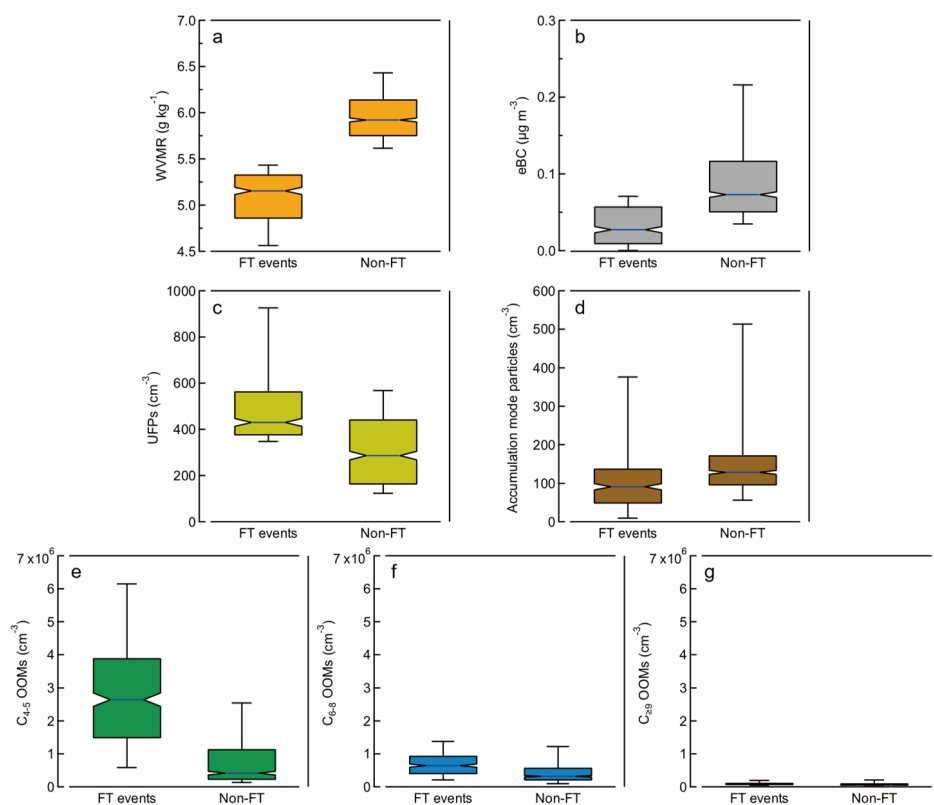

**Figure 2. Variation in the parameters measured during periods with FT events and without FT events (non-FT).** Only nighttime data during the study period (January 2018) are included. **a**, WVMR concentrations. **b**, eBC mass concentrations. **c**, Number concentrations of UFPs. **d**, Number concentrations of accumulation mode particles. **e**, Concentrations of $C_{4-5}$ OOMs. **f**, Concentrations of $C_{6-8}$ OOMs. **g**, Concentrations of $C_{\geq 9}$ OOMs. Concentrations of $C_{\geq 9}$ OOMs were close to the detection limit during the study period. OOMs with less than four carbon atoms (not shown here) were mostly composed of small organic acids, including malonic acid ($C_3H_4O_4$) and oxalic acid ($C_2H_2O_4$), which could come from both biogenic and anthropogenic sources. Boxes and whiskers are plotted for the 10th, 25th, 50th, 75th, and 90th percentiles. Notches denote the 95% confidence interval of the median value. The number of data points (10-minute resolution) for FT events and non-FT periods are 370 and 842, respectively.

We further compare the molecular composition of OOMs during FT, non-FT, and daytime events (Fig. 3). The variability in OOM composition among these periods strongly indicates different sources, pathways, and origins of OOMs observed at CHC. During FT events, $C_{4-5}$ OOMs such as $C_4H_{6,8}O_{4,5}$ and $C_5H_{6,8,10}O_{4,5}$ were the dominant CHO species (Fig. 3a). These compounds have been observed previously in chamber studies on the oxidation of isoprene ($C_5H_8$) by hydroxyl radicals (OH) (Krechmer et al., 2015; Wennberg et al., 2018). The most abundant CHONs (Fig. 3b) were $C_4H_{7,9}O_{3,4}(ONO_2)$ and $C_5H_{7,9}O_{3,4}(ONO_2)$, which have been characterized as OH oxidation products of isoprene in the presence of $NO_x$ (Lee et al., 2016; Massoli et al., 2018; Wennberg et al., 2018). $C_{6-8}$ OOMs observed in FT events consisted of several compounds, mainly $C_6H_{10}O_6$, $C_7H_{9,11}O_5(ONO_2)$, and $C_8H_{12,14}O_6$, which could be from various sources, such as the residual of daytime BL air, evaporation from existing particles, and oxidation products of other BVOCs (e.g., monoterpenes). During non-FT periods (Fig. 3c, d), OOM concentrations were generally lower. The $C_{4-5}$ OOM composition was very similar to those identified in FT events, but almost an order of magnitude lower in concentration. In contrast, $C_{6-8}$ OOMs were more varied and accounted for a larger fraction of the total OOMs than during FT events. Whereas we attribute the significantly reduced $C_{4-5}$ OOM concentrations to the decreased influence of FT air, the changes in $C_{6-8}$ OOMs appear to be due to the increased impact of BL air (or nighttime residual BL air). During non-FT events, $C_{4-5}$ OOMs during daytime were accompanied by noticeable amounts of $C_{6-8}$ and $C_{\geq 9}$ OOMs (Fig. 3e, f). These OOMs also contained on average more oxygen atoms than the OOMs in FT air, suggesting they had a different origin, likely oxidation products of urban biogenic and anthropogenic emissions in the nearby metropolitan area (Massoli et al., 2018).

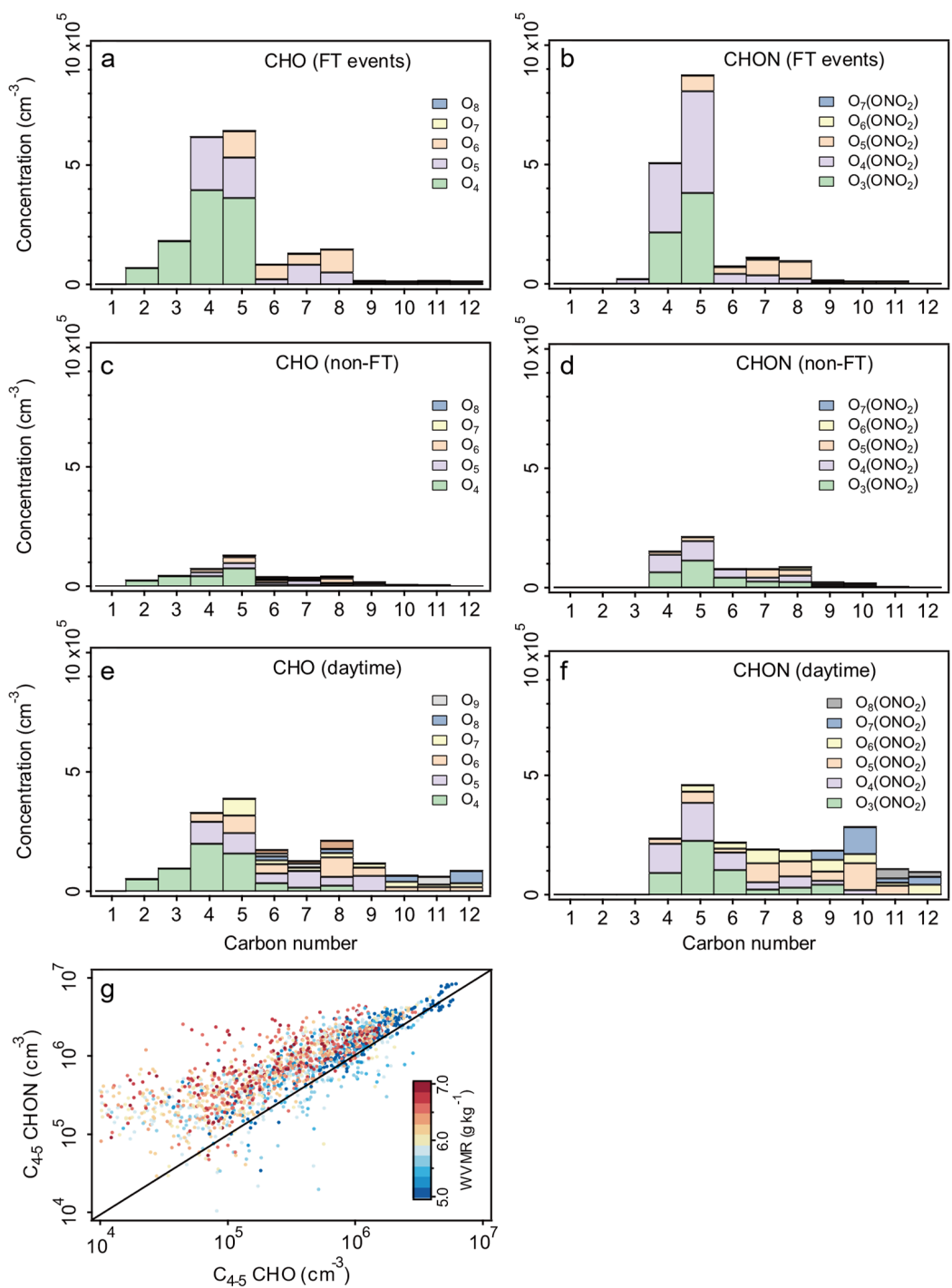

**Figure 3. Chemical composition of OOMs observed in different conditions at CHC in January 2018. a** and **b**, Concentrations of OOMs containing (**a**) carbon, hydrogen, and oxygen atoms (CHO) and (**b**) carbon, hydrogen, oxygen, and nitrogen atoms (CHON) averaged over all FT events. OOMs are grouped with different oxygen atom numbers as a function of carbon atom numbers. Note that the contribution of a nitrate (-ONO$_2$) functional group to OOM volatility is comparable to an alcohol (-OH) group due to the same effective O:C ratio[29]. **c** and **d**, Concentrations of (**c**) CHO and (**d**) CHON OOMs averaged over "non-FT" periods during nighttime (19:00–06:00). **e** and **f**, Concentrations of (**e**) CHO and (**f**) CHON OOMs averaged over daytime (07:00 - 18:00) when CHC was affected by BL air from the nearby La Paz – El Alto metropolitan area. **g**, Correlation between C$_{4-5}$ CHO and C$_{4-5}$ CHON in January 2018 (including

both daytime and nighttime data), colored by the WVMR. Here the FT events are indicated by lower WVMR and higher OOM concentrations.

We believe that the dominant role of $C_{4-5}$ compounds in the observed OOMs at CHC during FT events directly reflects strong isoprene emissions from the Amazonian rainforest. During the wet season, isoprene concentrations (several ppbv (parts per billion by volume)) in the Amazon BL are approximately an order of magnitude higher than those of other BVOCs (e.g., monoterpenes ($C_{10}H_{16}$)) (Kesselmeier et al., 2000). Shilling *et al.* reported that isoprene was still abundant (up to ~3 ppbv) at the top of the Amazon BL, whereas monoterpene concentrations were usually below 0.2 ppbv (the instrument detection limit) (Shilling et al., 2018). Frequent mesoscale convective systems in the Amazon Basin (Rehbein et al., 2018) provide a transport mechanism for lifting organics from the BL to the tropical FT (Andreae et al., 2018; Kulmala et al., 2006). In this process, while organics less volatile and/or more soluble than isoprene are largely scavenged by cloud hydrometeors (Bardakov et al., 2021), a substantial amount of isoprene (>1 ppbv, originating from the Amazon BL) is transported to the FT over tropical South America (Palmer et al., 2022). The outflow of these mesoscale convective systems in the FT provides a suitable chemical environment for the formation of isoprene-derived OOMs (isoprene-OOMs) observed in our study. During the study period, intensive lightning activity associated with convective systems was observed (Fig. S3 and S9), which enriches the amounts of OH and nitric oxide (NO) needed for forming isoprene-OOMs in this region and the Amazon FT (Palmer et al., 2022; Brune et al., 2021). As a result, the overall CHON/CHO ratio of isoprene-OOMs observed in FT events at CHC was ~1.3:1 (Fig. 3g). A similar ratio was determined in an isoprene-dominated and low-NOx ($NO_x$ = NO + $NO_2$) environment at a ground forest site in Alabama, United States, where OOMs derived from isoprene were also much higher than those derived from monoterpene (Massoli et al., 2018). OOMs previously condensed on aerosol particles in the upper part of FT can be another potential source of these isoprene-OOM vapors, due to evaporation as temperature rises in descending air masses (Jonsson et al., 2007). Thus, we conclude that the majority of OOMs observed in lower tropical FT at CHC were from the isoprene emitted from the Amazonian rainforest. Still, it is important to note that the chemical environment changes along the path of the air masses, and the chemical transformation of OOMs may happen in the gas and particle phases. However, a detailed investigation of this mechanism is unlikely based on our mountain-top measurements.

**Discussion**

The question is whether the observed isoprene-OOMs play a role in NPF to form large numbers of UFPs observed over the tropical FT (Fig. 4) (Williamson et al., 2019; Wang et al., 2016; Andreae et al., 2018). Oxidation products of isoprene tend to be more volatile than other BVOCs with higher carbon numbers, such as monoterpenes (Heinritzi et al., 2020). They show a suppressive effect on NPF in laboratory studies (Kiendler-Scharr et al., 2009; McFiggans et al., 2019), and are likely the reason for NPF to be rarely observed in the Amazon BL (Wang et al., 2016; Andreae et al., 2018). As a result of their high volatility, these isoprene-OOMs can undergo long-range transport in the lower and middle regions of tropical FT (~2 to 9 km a.s.l.; see (Andreae et al., 2018) and Methods) and have a spatially broader influence on FT aerosol compared to non-isoprene OOMs. A recent study by Palmer *et al.* showed that isoprene epoxydiols (IEPOX) in the FT possibly enhanced organic aerosol formation in the tropics (Palmer et al., 2022). Isoprene-OOMs, another isoprene oxidation pathway, by contrast, may participate in NPF more efficiently, as they are less volatile than IEPOX under similar FT conditions due to the higher oxidation states (see (Donahue et al., 2012) and Methods).

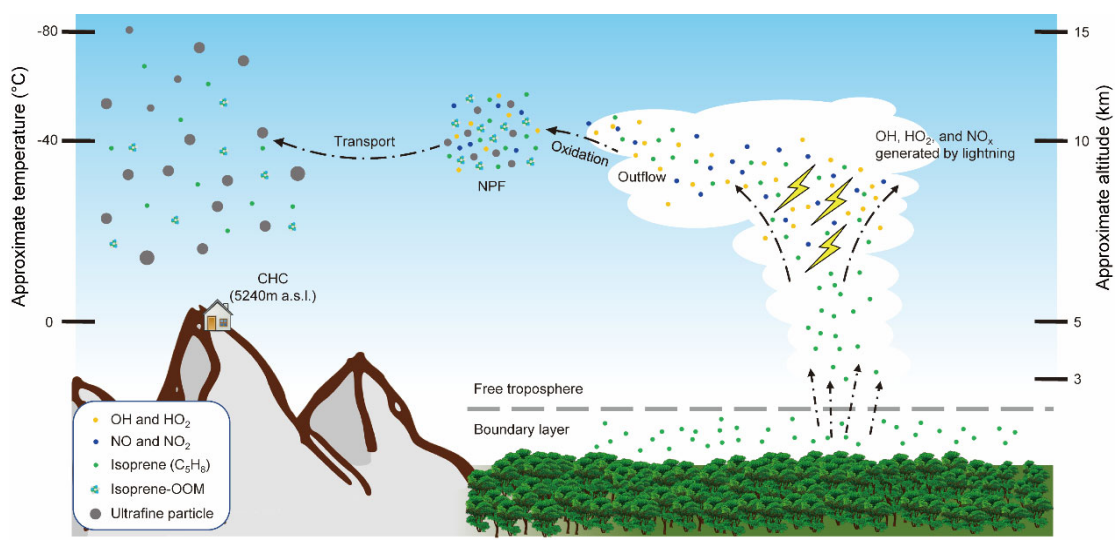

**Figure 4. Life cycle of isoprene-OOMs and their role in NPF over the tropical FT.** Right, isoprene emitted from the Amazon rainforest is transported upwards by mesoscale convective systems (convective clouds) to the tropical FT. Isoprene surviving from removal processes in the clouds (e.g., scavenging by cloud hydrometeors) reacts with hydroxyl radicals (OH), hydroperoxyl radicals (HO$_2$), and nitrogen oxides (NO and NO$_2$) produced by lightning in the convective clouds (Brune et al., 2021). In the cloud outflow, isoprene-OOMs (C$_{4-5}$ CHOs and C$_{4-5}$ CHONs) are formed. These OOMs are potentially important contributors to particle growth (middle). They can also endure long-range transport in the lower and middle regions of tropical FT and contribute to aerosol formation over a large scale in the tropics (left).

In addition to the gas phase observations during the wet season, enhanced contributions of $C_{4-5}$ OOMs were observed in the particle phase at CHC when particle chemical composition measurements were available (in April 2018 during the wet-to-dry transition period; Methods). On the night of 22 April 2018, CHC was affected by FT air from the western part of Amazonia (Fig. S10). As in the winter FT events, concentrations of gaseous $C_{4-5}$ OOMs increased with decreasing WVMR and eBC concentration, while $C_{6-8}$ and $C_{\geq 9}$ OOMs decayed during that time (from 19:00 to 21:00). Particulate $C_{4-5}$ OOMs mass fraction simultaneously increased, indicating that isoprene-OOMs were involved in particle growth in the Amazon FT.

We present a comprehensive analysis of organic trace gases and atmospheric aerosol in-situ observations at CHC in Bolivia, one of the highest atmospheric observatories in the world. Mountain-top state-of-the-art mass spectrometry, combined with detailed, highly resolved air mass analyses demonstrates the presence of OOMs in both gas and particle phases in tropical FT air from Amazonia. Our results indicate that the observed OOMs are dominated by oxidation products of isoprene emitted from the rainforest more than 800 kilometers away, and thus likely contribute to the growth stage of the NPF under the tropical FT condition on a continental scale. Such molecular-level OOMs measurements are unprecedented for the tropical FT, and are crucial for understanding aerosol formation over the tropics.

**Methods**

*CHC station*

The Global Atmosphere Watch (GAW) station Chacaltaya (CHC, 16.3505 S, 68.1314 W, Extended Data Fig. 1) is located at 5240 m above sea level (a.s.l.) near the summit of Mount Chacaltaya in the Bolivian Andes. It is ~17 km north and ~1.6 km above the La Paz – El Alto metropolitan area, which has a population of ~1.7 million, and close to the Bolivian Amazonia (including Beni, Santa Cruz, north of La Paz departments). A detailed description of the station and its surrounding area can be found on the website of the CHC station (http://www.chacaltaya.edu.bo/) and in the studies conducted at CHC (Bianchi et al., 2021; Chauvigne et al., 2019; Wiedensohler et al., 2018).

*SALTENA campaign*

The Southern hemisphere high ALTitude Experiment on particle Nucleation And growth (SALTENA) campaign was conducted from December 2017 to May 2018, aiming to understand the formation/growth mechanism and properties of aerosols measured at CHC. The six-month measurement campaign was arranged in order to cover the wet season (December to ~Feburary), transition period (~March to April), and dry season (May). Our study mainly focuses on measurements during the wet season in January 2018, when CHC was significantly affected by Amazon FT air (Aliaga et al., 2021) and OOM measurement data were continuously available (i.e., from 6 to 22 January 2018).

**Instrumentation**

*APi-TOF*

The Atmospheric pressure interface time-of-flight mass spectrometer (APi-TOF, TOFWERK AG and Aerodyne Research) was deployed to measure the chemical composition of naturally charged negative ions in January 2018 at CHC. The atmospheric pressure interface (APi) allows the instrument to sample ions in ambient air directly by reducing the pressure of the sampled airflow (14 standard liters per minute, SLPM, in total and 0.8 SLPM go into the instrument) from atmospheric pressure to ~ $10^{-4}$ mbar. The ions are focused and guided by two quadrupoles and an ion lens in the APi before entering the time-of-flight mass spectrometer (TOF-MS; ~$10^{-6}$ mbar). In this part, ions are detected and identified. The resolving power was ~ 5000

Th/Th. A description with more details of this instrument is presented in Junninen *et al.* (2010). APi-TOF data used in this study were averaged to 1-hour resolution.

*CI-APi-TOF*

The nitrate-based chemical ionization atmospheric pressure interface time-of-flight mass spectrometer (CI-APi-TOF, TOFWERK AG and Aerodyne Research) is an APi-TOF coupled with a chemical ionization unit (CI) using nitric acid ($HNO_3$) as the ionization reagent. The instrument is extensively used to measure oxidized organic compounds and sulfuric acid ($H_2SO_4$; Jokinen et al., 2012). In the CI module, the nitrate ion is generated by exposing the sheath flow (20 SLPM) that contains $HNO_3$ to soft x-ray radiation, and then charging the neutral molecules in the sampling airflow (10 SLPM) within a reaction time of ~200 ms before they enter the APi and the TOF-MS modules. The instrument was calibrated with $H_2SO_4$ using the setup described in Jokinen et al.[32]. After including the diffusion loss of $H_2SO_4$ in the 1.5 m sampling line, a calibration factor of $1.5 \times 10^{10}$ molecules cm$^{-3}$ was obtained and used for determining the concentration of oxidized organic molecules (OOMs).

The OOMs concentrations were estimated in two steps. We first corrected the measured signal intensity with a mass-dependent transmission function which depends on the setting of the instrument and was determined by using the method described in a previous study (Dunne et al., 2016). After the transmission correction, the calibration factor of $H_2SO_4$ was used to estimate the observed OOMs concentrations. It is important to note that OOMs might not be charged as efficiently as $H_2SO_4$ by nitrate ion ($NO_3^-$) in the CI unit (Hyttinen et al., 2015). As a result, OOMs concentrations presented in this study could be underestimated. A lower detection limit of $\sim 5 \times 10^4$ molecule cm$^{-3}$ was determined from $H_2SO_4$ and zero measurements. The CI-APi-TOF data included in this study were averaged to the 10-minute resolution.

*FIGAERO HR-TOF-CIMS*

The filter inlet for gases and aerosols (FIGAERO) coupled to a high-resolution time-of-flight chemical ionization mass spectrometer (FIGAERO HR-TOF-CIMS, Aerodyne Research) using iodide ($I^-$) as the reagent ion was deployed to measure the molecular composition of organic compounds and inorganic acids. The FIGAERO inlet can be operated in gas-phase and particle-phase modes. In the gas-phase mode, ambient air is directly sampled into the ion-molecule reactor while particles are simultaneously collected to a polytetrafluoroethylene filter through another sampling port. In the particle-phase mode, a nitrogen gas stream is heated and blown through the filter to evaporate the particles via temperature-programmed desorption. More details about the instrument can be found in Mohr et al. (2019). It is important to note that the FIGAERO HR-TOF-CIMS was only deployed from April (transition season) at CHC, and no measurement was available in January 2018 during our study period (wet season).

*MPSS*

The mobility particle size spectrometer (MPSS) was deployed to measure particle number size distribution in a size range of 10 to 500 nm at CHC (Wiedensohler et al., 2012). The instrument consists of a bipolar diffusion charger, a Hauke-type differential mobility analyzer, and a TSI 3772 condensation particle counter.

*MAAP*

The Multi-Angle Absorption Photometer (MAAP, Thermo-Scientific model 5012) was used to determine equivalent black carbon (eBC) mass concentrations at CHC (Petzold et al., 2005). The lower detection limit of this instrument is ~ 0.05 μg m$^{-3}$ at a time resolution of 10 min.

*Ancillary measurements*

Air temperature, relative humidity (with respect to water), and atmospheric pressure were measured with an automatic weather station (AWS) at CHC.

**Model simulation**

*WRF simulation*

The weather research and forecasting model (WRF) is an advanced non-hydrostatic numerical weather prediction model (Skamarock et al., 2008) that can reproduce meteorological situations at a wide range of spatial and temporal scales. In this study, WRF version 4.0.3 was used. Four nested domains were used with grid spacings ranging from 38 km (the outmost domain, with the Amazon Basin, tropical Andes, and west Pacific ocean included) to 1 km (the innermost domain, including complex mountainous topography surrounding CHC and the interface between the Andes and the Amazon Basin included). A detailed description of the model setup and parameterization is given in Aliaga et al., (2021).

*FLEXPART-WRF simulation*

The 96-hour history and footprint of air masses arriving at CHC were determined using the flexible particle dispersion model (FLEXPART). Different versions of the FLEXPART model have been developed to adapt to a range of numerical weather prediction models. In this study, we used the latest version of FLEXPART-WRF model (version 3.3.2; Brioude et al., 2013) driven by the high-resolution meteorological output from the WRF simulation. In this way, more validity and accuracy were added to the FLEXPART air mass history simulation and made this simulation state-of-the-art compared to other similar studies in the Southern Hemisphere tropics of South America.

In the FLEXPART simulations, we continuously release 20,000 particles per hour from the 0-10 m layer above ground level (a.g.l.) at CHC. The output of FLEXPART-WRF, when running in backward mode, is a source receptor relationship (SRR). The SRR is the aggregated residence time of the passive air tracer particles at each three-dimensional (3-D) grid cell. The values are normalized by the total number of particles such that if all of the 20 thousand particles were to reside in only one cell, the SRR values of this cell would equal 96h (i.e., the total backward simulation time). The derived SRRs describe the relationship between each 3-D grid cell in the simulation (as potential source regions) and the air masses arriving at the measurement station (receptor). In general, a higher SRR of a grid cell indicates a larger contribution to the observed air masses. The detailed model setup is also described in Aliaga et al. (2021).

**Estimation of the WVMR**

The water vapor mixing ratio (WVMR) is a measure of the mass concentration of water vapor in the atmosphere, which is calculated as follows:

$$WVMR = B \times \frac{P_w}{P_{tot} - P_w} \qquad (1)$$

where B is a constant (621.9907 g kg$^{-1}$, molecular weight ratio of water to dry air); $P_w$ and $P_{tot}$ are water vapor pressure and the atmospheric pressure, respectively. In this study, $P_w$ was determined based on the method presented in Buck et al. (1981), using the ambient temperature, RH, and pressure measured at CHC.

**Identification of FT events**

The periods when free troposphere (FT) air dominated at CHC were identified as FT events, as mentioned in the main text. Air mass history analysis shows that, on average, FT air constituted ~70% of air masses arriving at CHC in January 2018. However, the observed air masses were usually not purely from the FT due to the concurrent local influence on CHC, such as boundary layer (BL) air from the nearby La Paz – El Alto metropolitan area. The impact of BL air at CHC increases with the effect of thermally-driven local air circulation during daytime (07:00 - 18:00; all the times are in local time, UTC - 4). This is indicated by the advecting eBC concentration at noontime at CHC (Wiedensohler et al., 2018). Therefore, to minimize the influence of the BL on CHC, we only used the measurements during nighttime (19:00 – 06:00; Chauvigne et al., 2019).

The presence of water in the atmosphere (measured by, e.g., WVMR and RH) and black carbon were used to identify the influence of FT air in the previous studies (Schmeissner et al., 2011). This is due to FT air in the tropics typically containing much less water and is much cleaner (biomass burning is inhibited by high-level rainfall and moisture during the wet season; Machado et al., 2018) than BL air (Ryoo et al., 2009). Thus, to characterize the influence of FT air, WVMR and eBC concentrations were introduced as further parameters to identify FT events at CHC. Thresholds of 5.5 g kg$^{-1}$ in the WVMR (denotes the lower 30 % of the nighttime WVMR during the study period) and 0.08 μg m$^{-3}$ in eBC concentration (eBC nighttime mean concentration during the study period) were used.

**Supplementary information**

Tables S1 and Figures S1–S10.

**Acknowledgment**

We thank the Bolivian staff of the IIF-UMSA (Physics Research Institute, UMSA) who work at CHC for their valuable work under difficult conditions, and the long-term observations performed within the framework of GAW and ACTRIS; the IRD (Institut de Recherche pour le Développement) personnel for the logistic and financial support during all the campaign including shipping and customs concerns; the CSC-IT Center for Science, Finland, for generous computational resources that enabled the WRF and FLEXPART-WRF simulations to be conducted; the use of imagery provided by services from NASA's Global Imagery Browse Services (GIBS), part of NASA's Earth Observing System Data and Information System (EOSDIS).

**Funding:** This research has received support from European Union (EU) H2020 program via the findings European Research Council (ERC; project CHAPAs no. 850614 and ATM-GTP no. 742206), the Marie Skłodowska Curie (CLOUD-MOTION no. 764991), the Finnish Centre of Excellence as well as the Academy of Finland (project no. 311932, 315203 and 337549), and the Knut and Alice Wallenberg Foundation (WAF project CLOUDFORM no. 2017.0165). P. Artaxo acknowledge funds from FAPESP – Fundação de Amparo à Pesquisa do Estado de São Paulo, grant 2017/17047-0.

**Author contributions:** Q. Z., and F. B. designed the research. Q. Z., D. A., C. W., W. S., L. H., E. P., Y. G., W. H., M. L., J. E., O. P., X. C., A. M., F. V., I. M., C. M., and F. B. conducted the measurements. Q. Z., D. A., R. K., V. S., D. W., C. M., and F. B. analyzed the data. Q. Z., wrote the manuscript with major input from D. A., R. K., V. S., C. M., D. W., and F. B., and further contributions from all other authors.

**Conflict of interests:** The authors declare no competing interests.

**References:**

Williamson CJ, Kupc A and Axisa D *et al.* A large source of cloud condensation nuclei from new particle formation in the tropics. *Nature* 2019; **574**: 399–403.

Weigel R, Borrmann S and Kazil J *et al.* In situ observations of new particle formation in the tropical upper troposphere: the role of clouds and the nucleation mechanism. *Atmos Chem Phys* 2011; **11**: 9983–10010.

Wang J, Krejci R and Giangrande S *et al.* Amazon boundary layer aerosol concentration sustained by vertical transport during rainfall. *Nature* 2016; **539**: 416–419.

Andreae MO, Afchine A and Albrecht R *et al.* Aerosol characteristics and particle production in the upper troposphere over the Amazon Basin. *Atmos Chem Phys 2018;* **18**: 921–961.

Kulmala M, Reissell A and Sipilä M *et al.* Deep convective clouds as aerosol production engines: Role of insoluble organics. *J Geophys Res* 2006; **111**: D17202.

Palmer PI, Marvin MR and Siddans R *et al.* Nocturnal survival of isoprene linked to formation of upper tropospheric organic aerosol, *Science* 2022; **375**: 562-566.

Zhao B, Shrivastava M and Donahue NM *et al.* High concentration of ultrafine particles in the Amazon free troposphere produced by organic new particle formation. *Proc Natl Acad Sci USA* 2020; **117**: 25344-25351.

Bianchi F, Sinclair VA and Aliaga D *et al.* The SALTENA experiment: Comprehensive observations of aerosol sources, formation and processes in the South American Andes. *Bull Am Meteorol Soc* 2021; **1**: 1–46.

Aliaga D, Sinclair VA and Andrade M *et al.* Identifying source regions of air masses sampled at the tropical high-altitude site of Chacaltaya using WRF-FLEXPART and cluster analysis. *Atmos Chem Phys* 2021; **21**: 16453–16477.

Wiedensohler A, Andrade M and Weinhold K *et al.* Black carbon emission and transport mechanisms to the free troposphere at the La Paz/El Alto (Bolivia) metropolitan area based on the Day of Census (2012). *Atmos Environ* 2018; **194**: 158–169.

Chauvigne A, Aliaga D and Sellegri K *et al.* Biomass burning and urban emission impacts in the Andes Cordillera region based on in situ measurements from the Chacaltaya observatory, Bolivia (5240a.s.l.). *Atmos Chem Phys* 2019; **19**: 14805–14824.

Sun DZ and Lindzen RS. Distribution of tropical tropospheric water vapor. *J Atmos Sci* 1993; **50**: 1643–1660.

Bianchi F, Tröstl J and Junninen H et al. New particle formation in the free troposphere: A question of chemistry and timing. *Science* 2016; **352**: 1109–1112.

Bianchi F, Garmash O and He X *et al.* The role of highly oxygenated molecules (HOMs) in determining the composition of ambient ions in the boreal forest. *Atmos Chem Phys* 2017; **17**: 13819–13831.

Krechmer JE, Coggon MM and Massoli P *et al.* Formation of Low Volatility Organic Compounds and Secondary Organic Aerosol from Isoprene Hydroxyhydroperoxide Low-NO Oxidation. *Environ Sci Technol* 2015; **49**: 10330–10339.

Wennberg PO, Bates KH and Crounse JD *et al.* Gas-Phase Reactions of Isoprene and Its Major Oxidation Products. *Chem Rev* 2018; **118**: 3337-3390.

Lee BH, Mohr C and Lopez-Hilfiker FD *et al.* Highly functionalized organic nitrates in the southeast United States: Contribution to secondary organic aerosol and reactive nitrogen budgets. *Proc Natl Acad Sci USA* 2016; **113**: 1516–1521.

Massoli P, Stark H and Canagaratna MR *et al.* Ambient Measurements of Highly Oxidized Gas-Phase Molecules during the Southern Oxidant and Aerosol Study (SOAS) 2013. *ACS Earth Space Chem* 2018; **2**: 653-672.

Kesselmeier J, Kuhn U and Wolf A *et al.* Atmospheric volatile organic compounds (VOC) at a remote tropical forest site in central Amazonia. *Atmos Environ* 2000; **34**: 4063–4072.

Shilling JE, Pekour MS and Fortner EC *et al.* Aircraft observations of the chemical composition and aging of aerosol in the Manaus urban plume during GoAmazon 2014/5. *Atmos Chem Phys* 2018; **18**: 10773–10797.

Rehbein A, Ambrizzi T and Mechoso CR. Mesoscale convective systems over the Amazon basin. Part I: climatological aspects. *Int J Climatol* 2018; **38**: 215–229.

Bardakov R, Thornton JA and Riipinen I *et al.* Transport and chemistry of isoprene and its oxidation products in deep convective clouds. *Tellus B Chem Phys Meteorol* 2021; **73**: 1–21.

Brune WH, McFarland PJ and Bruning E *et al.* Extreme oxidant amounts produced by lightning in storm clouds. *Science* 2021; **372**: 711-715.

Jonsson ÅM, Hallquist M and Saathoff H. Volatility of secondary organic aerosols from the ozone initiated oxidation of -pinene and limonene. *J Aerosol Sci* 2007; **38**: 843–852.

Heinritzi M, Dada L and Simon M *et al.* Molecular understanding of the suppression of new-particle formation by isoprene. *Atmos Chem Phys* 2020; **20**: 11809–11821.

Kiendler-Scharr A, Wildt J and Maso MD *et al.* New particle formation in forests inhibited by isoprene emissions. *Nature* 2009; **461**: 381–384.

McFiggans G, Mentel TF and Wildt JJ *et al.* Secondary organic aerosol reduced by mixture of atmospheric vapours. *Nature* 2019; **565**: 587–593.

Donahue NM, Kroll JH and Pandis SN et al. A two-dimensional volatility basis set – Part 2: Diagnostics of organic-aerosol evolution. *Atmos Chem Phys* 2012; **12**: 615–634.

Chuang WK and Donahue NM. A two-dimensional volatility basis set – Part 3: Prognostic modeling and NOx dependence. *Atmos Chem Phys* 2016; **16**: 123–134.

Rose C, Sellegri K, and Asmi E *et al.* Major contribution of neutral clusters to new particle formation at the interface between the boundary layer and the free troposphere. *Atmos Chem Phys* 2015; **15**: 3413–3428.

Junninen H, Ehn M and Petäjä T *et al.* A high-resolution mass spectrometer to measure atmospheric ion composition. *Atmos Meas Tech* 2010; **3**: 1039–1053.

Jokinen T, Sipilä M and Junninen H *et al.* Atmospheric sulphuric acid and neutral cluster measurements using CI-APi-TOF. *Atmos Chem Phys* 2012; **12**: 4117–4125.

Dunne EM, Gordon H and Kürten A *et al.* Global atmospheric particle formation from CERN CLOUD measurements. *Science* 2016; **354**: 1119–1124.

Hyttinen N, Kupiainen-Määttä O and Rissanen MP *et al.* Modeling the Charging of Highly Oxidized Cyclohexene Ozonolysis Products Using Nitrate-Based Chemical Ionization. *J Phys Chem A* 2015; **119**: 6339–6345.

Mohr C, Thornton JA and Heitto A *et al.* Molecular identification of organic vapors driving atmospheric nanoparticle growth. *Nat Commun* 2019; **10**: 4442.

Wiedensohler A, Birmili W and Nowak A *et al.* Mobility particle size spectrometers: Harmonization of technical standards and data structure to facilitate high quality long-term observations of atmospheric particle number size distributions. *Atmos Meas Tech* 2012; **5**: 657–685.

Petzold A, Schloesser H and Sheridan PJ *et al.* Evaluation of Multiangle Absorption Photometry for Measuring Aerosol Light Absorption. *Aerosol Sci Technol* 2005; **39**: 40–51.

Skamarock WC, Klemp JB and Dudhi J *et al.* A Description of the Advanced Research WRF Version 3. Tech. Note No. NCAR/TN-475+STR, University Corporation for Atmospheric Research; 2008.

Brioude J, Arnold D and Stohl A *et al.* The Lagrangian particle dispersion model FLEXPART-WRF version 3.1. *Geosci Model Dev* 2013; **6**: 1889–1904.

Buck AL. New equations for computing vapour pressure and enhancement factor. *J Appl Meteorol* 1981; **20**: 1527–1532.

Schmeissner T, R. Krejci R and Ström J *et al.* Analysis of number size distributions of tropical free tropospheric aerosol particles observed at Pico Espejo (4765 m a.s.l.), Venezuela. *Atmos Chem Phys* 2011; **11**: 3319–3332.

Machado LAT, Calheiros AJP and Biscaro T *et al.* Overview: Precipitation characteristics and sensitivities to environmental conditions during GoAmazon2014/5 and ACRIDICON-CHUVA. *Atmos Chem Phys* 2018; **18**: 6461–6482

Ryoo JM, Igusa T and Waugh DW. PDFs of tropical tropospheric humidity: Measurements and theory. *J Clim* 2009; **22**: 3357–3373.

**Table S1** List of FT events observed in January 2018.

| Date | FT event | Start time | End time | Mean WVMR (g kg$^{-1}$) | Mean eBC (μg m$^{-3}$) | Note |
|---|---|---|---|---|---|---|
| **06/01/18** | No | | | | | |
| **07/01/18** | Yes | 01:40 | 05:00 | 5.23 | 0.07 | Precipitation from 02:20 to 03:30 |
| **08/01/18** | Yes | 23:10 | 03:40 (+1 day) | 5.28 | <0.05* | |
| **09/01/18** | Yes | 23:00 | 00:40 (+1 day) | 5.36 | 0.06 | |
| **10/01/18** | Yes | 02:30 | 07:00 | 5.13 | <0.05* | |
| | | 21:10 | 00:00 | 5.20 | <0.05* | |
| **11/01/18** | Yes | 19:10 | 23:20 | 5.31 | 0.06 | |
| **12/01/18** | Yes | 03:10 | 06:10 | 5.22 | <0.05* | |
| **13/01/18** | No | | | | | |
| **14/01/18** | Yes | 03:40 | 06:30 | 4.99 | 0.06 | |
| **15/01/18** | No | | | | | |
| **16/01/18** | No | | | | | |
| **17/01/18** | Yes | 02:40 | 06:00 | 4.91 | <0.05* | |
| | | 19:50 | 22:10 | 5.22 | <0.05* | |
| **18/01/18** | Yes | 01:10 | 06:00 | 5.32 | <0.05* | |
| | | 19:30 | 02:50 (+1 day) | 5.15 | <0.05* | |
| **19/01/18** | Yes | 19:20 | 21:40 | 5.35 | <0.05* | |
| **20/01/18** | Yes | 04:00 | 07:00 | 5.30 | <0.05* | |
| **21/01/18** | Yes | 20:30 | 22:30 | 5.28 | <0.05* | |
| **22/01/18** | Yes | 01:00 | 06:10 | 5.32 | <0.05* | |

*Below the threshold of MAAP (0.05 μg m$^{-3}$).

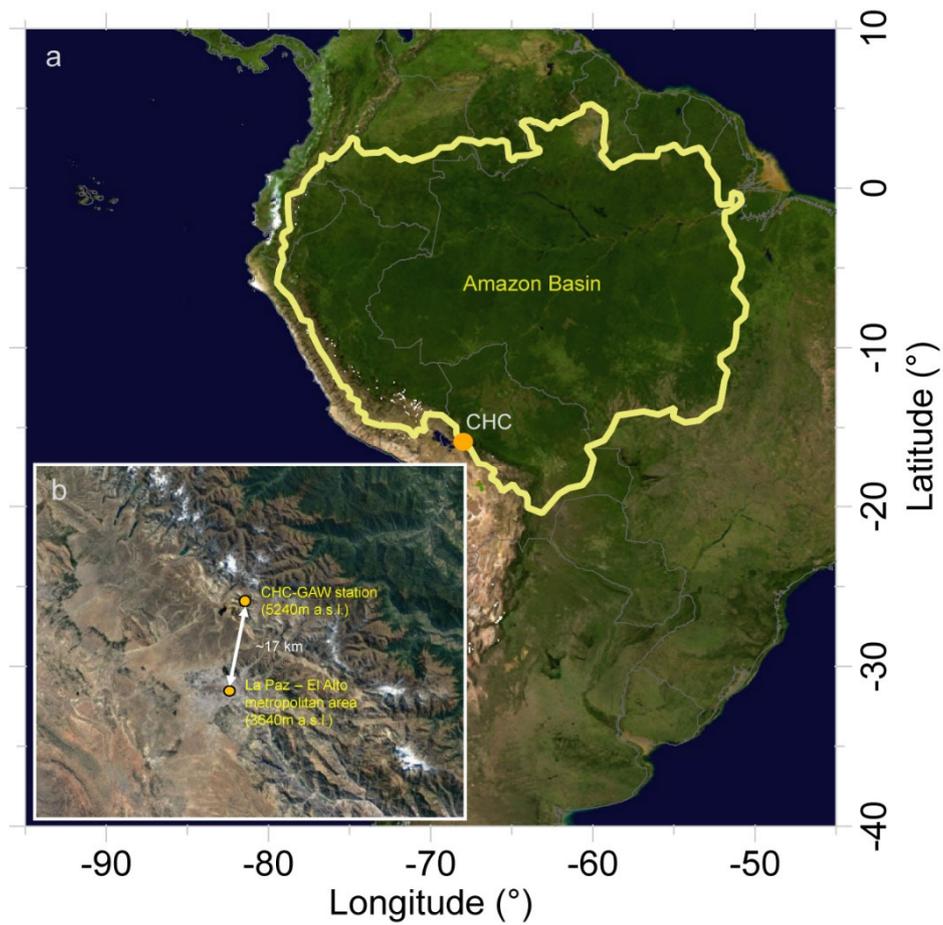

**Figure S1. a**, Google Earth satellite images showing locations of CHC, **b**, La Paz- El Alto metropolitan area. The yellow outline denotes the borders of the Amazon Basins (Charity et al., 2016).

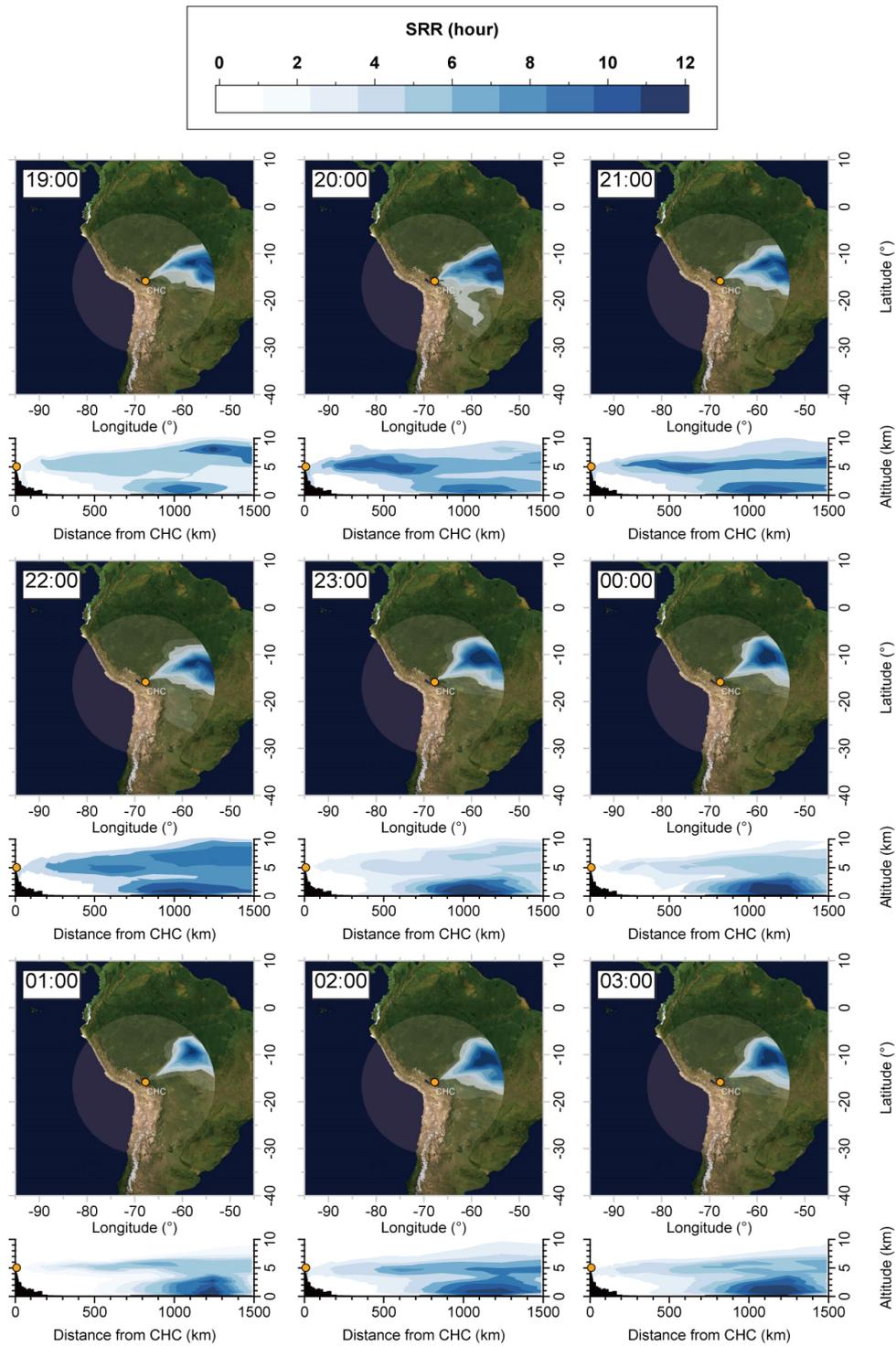

**Figure S2.** 96-hour WRF-FLEXPART air mass history and footprint for each hour from 19:00 – 03:00 on the night of 10 January 2018. The color bar denotes the SRR of the passive air tracers integrated in the radial direction. The black shaded area indicates the topography near the station. The lower SRR intensities in the region close to CHC are likely due to the short residence time of the passive air tracers. The semi-transparent circle indicates the horizontal output domain of the model.

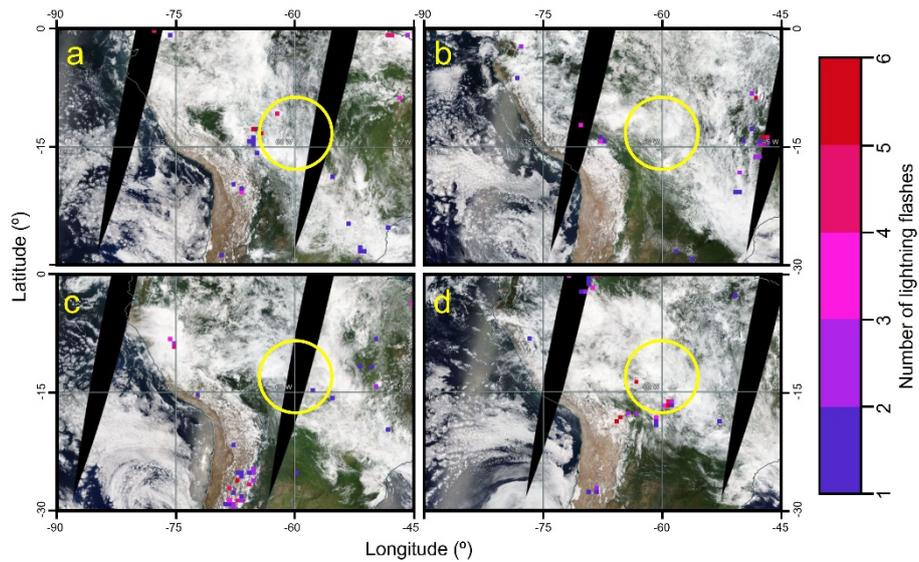

**Figure S3.** True-color daily satellite images from Terra/MODIS corrected reflectance imagery (Gumley et al., 2010) and daily lightning flash counts International Space Station (ISS) Lightning Imaging Sensor (LIS) data (Blakeslee et al., 2021) on **a**, 10 January 2018, **b**, 9 January 2018, **c**, 8 January 2018, and **d**, 7 January 2018. The yellow circle in each panel is a rough indication of the region where the SRRs have the highest intensities in Fig. 1.

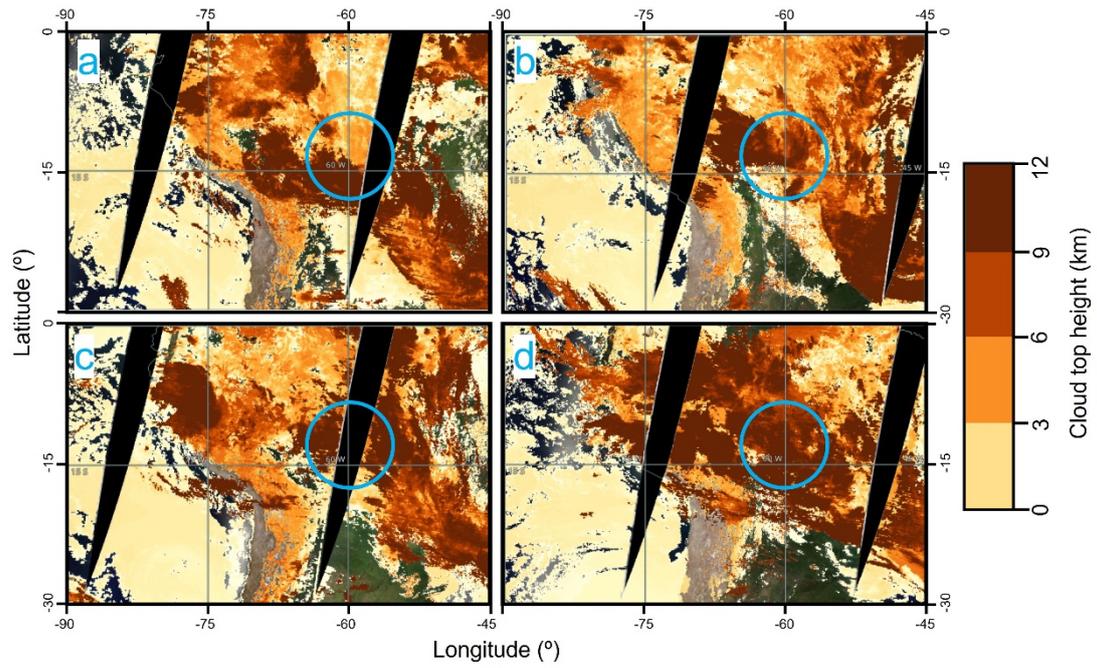

**Figure S4.** Cloud top height satellite images from Terra/MODIS on **a**, 10 January 2018, **b**, 9 January 2018, **c**, 8 January 2018, and **d**, 7 January 2018. The blue circle in each panel is a rough indication of the region where the SRRs have the highest intensities in Fig. 1.

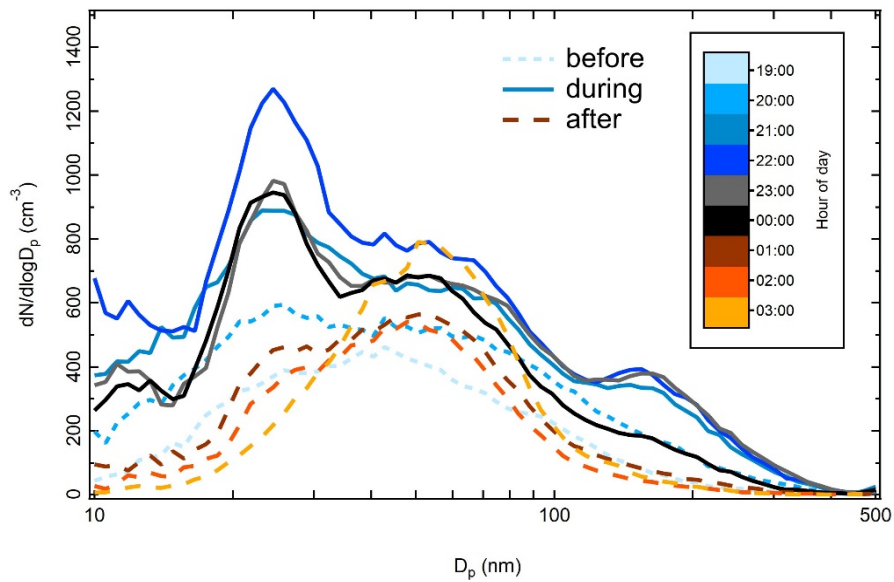

**Figure S5.** Evolution of the hourly averaged particle number size distribution during the night of 10 January 2018. Dashed, solid, and dotted lines denote particle size spectra observed before (19:00 – 20:00), during (21:00 – 00:00), and after (01:00 – 03:00) the FT event, respectively. It is noted that the influence of FT air on the observed aerosol particles at CHC is already evident at 20:00. The concurrent increases in accumulation mode particles indicated that a fraction of the UFPs originating in the Amazon FT could have grown to larger sizes during the transport to CHC (takes ~2 days; Froyd et al., 2009).

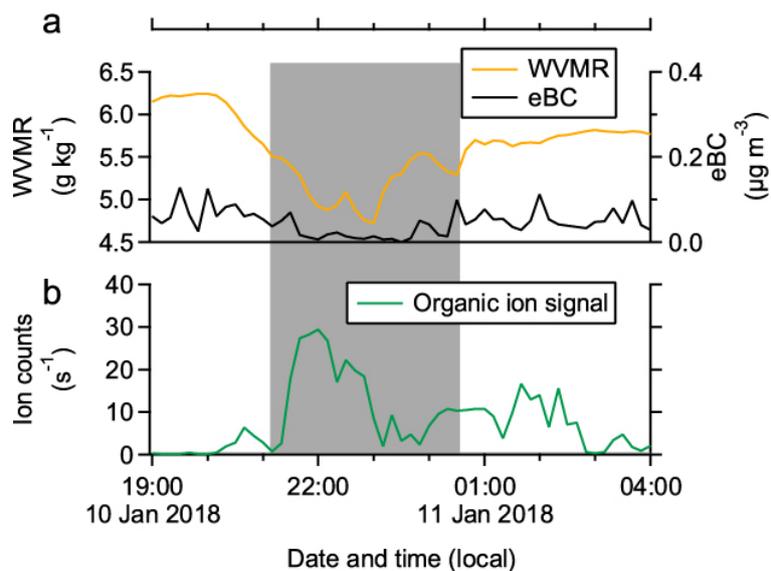

**Figure S6.** The Amazon FT event observed at CHC during the night of 10 January 2018 (as shown in Fig. 1). **a**, Concentrations of the water vapor mixing ratio (WVMR) and eBC at CHC. The grey shaded area denotes the FT event period (from ~21:00 to ~01:00, WVMR ≤5.5 g kg-1 and eBC ≤0.08 μg m-3) at CHC. **b**, Signal of organic ions measured with the APi-TOF.

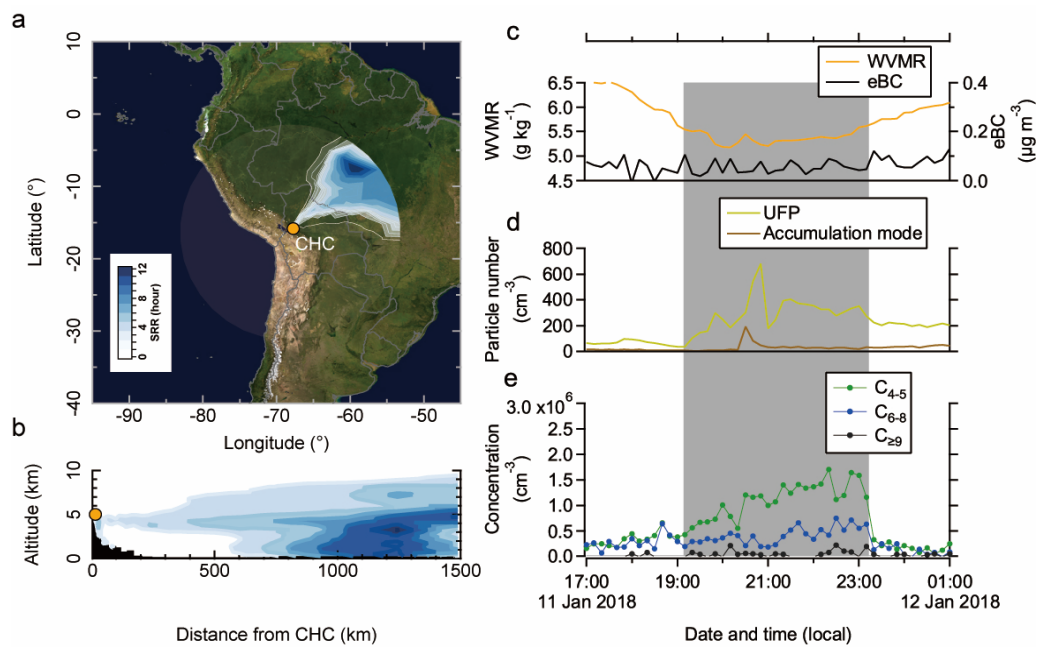

**Figure S7.** An Amazon FT event observed at CHC during the night of 11 January 2018. **a**, Map and horizontal profile of the vertically integrated SRR averaged from 19:00 to 23:00. The color bar denotes the SRR of the passive air tracers. **b**, Vertical profile of the SRR integrated in the radial direction and averaged from 19:00 to 23:00. The black shaded area indicates the topography condition near the station. **c**, Concentrations of WVMR and eBC. **d**, Concentrations of UFPs and accumulation mode particles. **e**, Concentrations of $C_{4-5}$, $C_{6-8}$, and $C_{\geq 9}$ OOMs. The grey shaded area denotes the FT event period (from 19:10 to 23:20) at CHC. Note that the scales of y-axis in the c, d, and e panels may be different from those in Fig. 1.

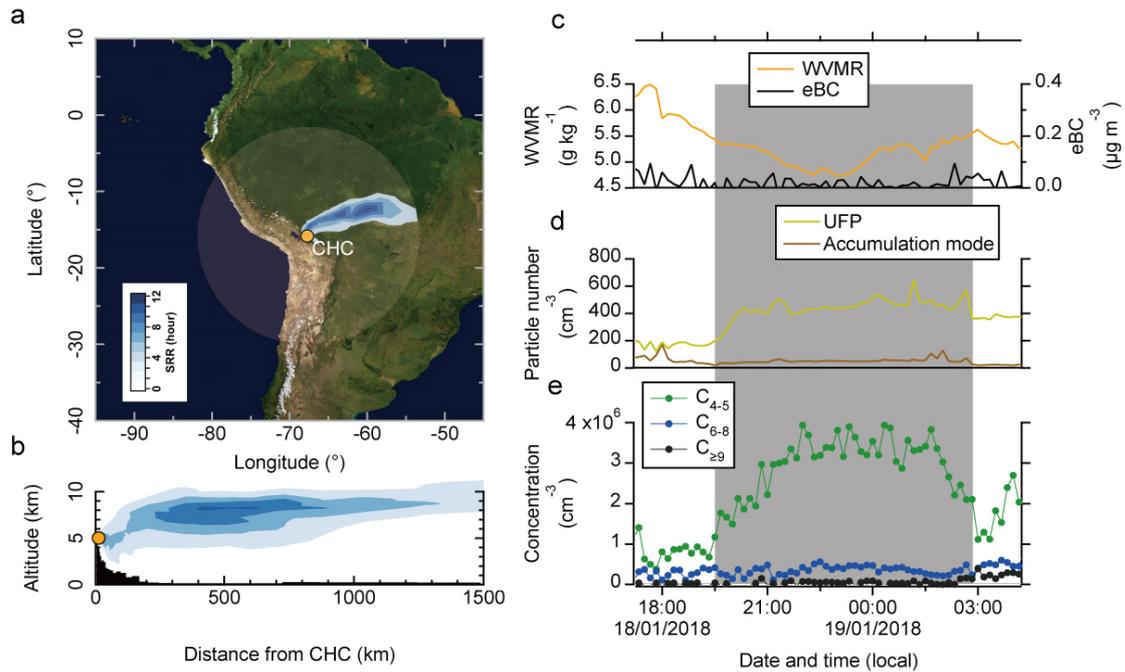

**Figure S8.** An Amazon FT event observed at CHC during the night of 18 January 2018. **a**, Map and horizontal profile of the vertically integrated SRR averaged from 20:00 to 03:00. The color bar denotes the SRR of the passive air tracers. **b**, Vertical profile of the SRR integrated in the radial direction and averaged from 20:00 to 03:00. The black shaded area indicates the topography condition near the station. **c**, Concentrations of WVMR and eBC. **d**, Concentrations of UFPs and accumulation mode particles. **e**, Concentrations of $C_{4-5}$, $C_{6-8}$, and $C_{\geq 9}$ OOMs. The grey shaded area denotes the FT event period (from 19:30 to 02:50) at CHC. Note that the scales of y-axis in the c, d, and e panels may be different from those in Fig. 1. The convection process in this FT event is not fully captured in the air mass history analysis, like-ly due to the lower strength and/or smaller scale compared to the events on 10 and 11 January 2018.

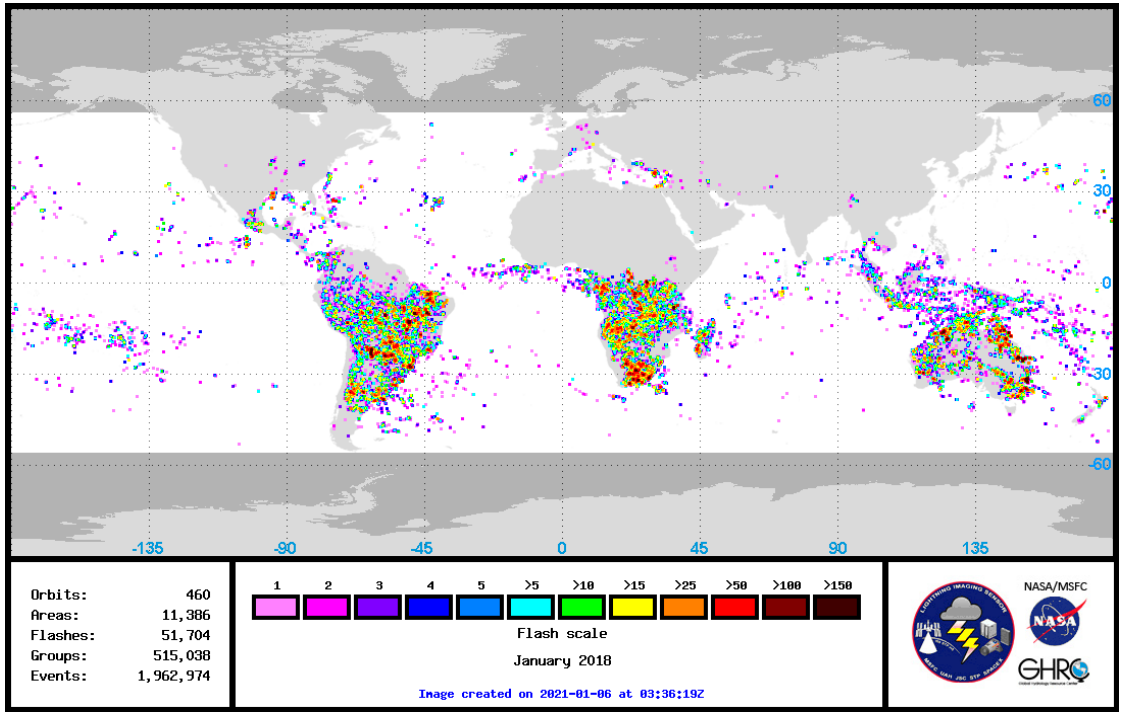

**Figure S9.** Global lightning activity distribution in January 2018 from the ISS LIS dataset (Andreae et al., 2018).

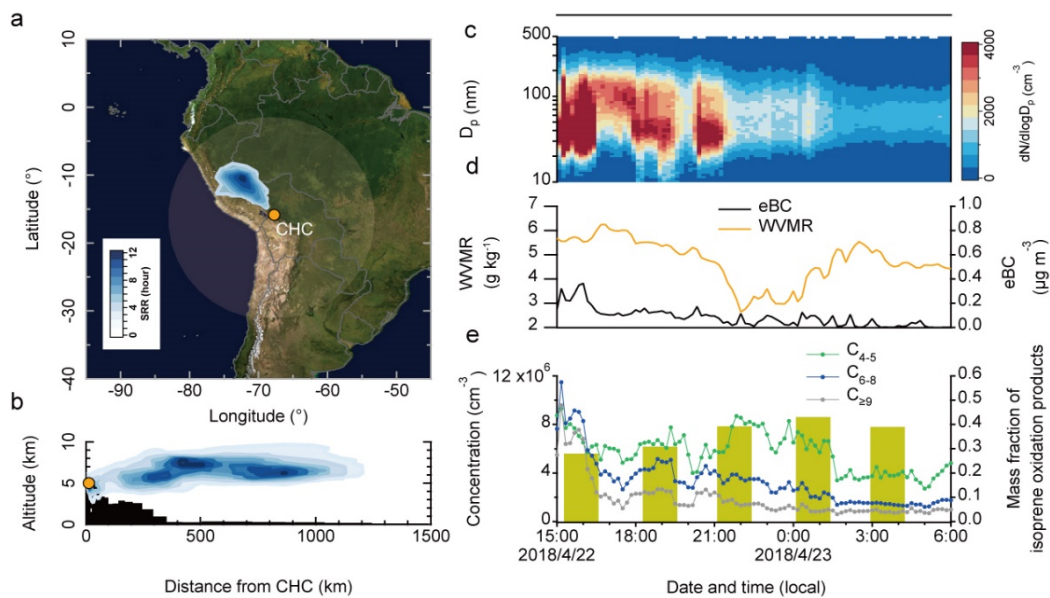

**Figure S10.** An Amazon FT event observed at CHC during the night of 22 April 2018. **a**, Map and horizontal profile of the vertically integrated SRR averaged from 21:00 to 02:00. **b**, Vertical profile of the SRR integrated in the radial direction and averaged from 21:00 to 02:00. The color bar denotes the SRR of the passive air tracers. The black shaded area in panel b denotes the topography condition near the station. **c**, Particle number size distribution (dN/dlogD$_p$) in 10-minute time resolution. **d**, Concentrations of eBC and the WVMR. **e**, Concentrations of C$_{4-5}$, C$_{6-8}$, and C$_{\geq 9}$ OOMs measured by the CI-APi-TOF and mass fraction of isoprene-derived OOMs (yellow bars) observed in FT air in the particle phase, measured by the FIGAERO HR-TOF-CIMS. The convection process in this FT event is not fully captured in the air mass history analysis, like-ly due to the lower strength and/or smaller scale compared to the events on 10 and 11 January 2018.

**Supplemental References**

Charity S, Dudley N and Oliveira D *et al.* Living Amazon Report 2016: A regional approach to conservation in the Amazon. WWF Living Amazon Initiative Brasília and Quito; 2016.

Gumley L. Creating Reprojected True Color MODIS Images: A Tutorial. Space Science Engineering Center University Wisconsin-Madison; 2010.

Blakeslee RJ. NRT Lightning Imaging Sensor (LIS) on International Space Station (ISS) Science Data (2021); doi: 10.5067/LIS/ISSLIS/DATA109.

Froyd KD, Murphy MD and Sanford TJ *et al.* Aerosol composition of the tropical upper troposphere. *Atmos Chem Phys* 2009; **9**: 4363–4385.

Andreae MO, Afchine A, and Albrecht R *et al.* Aerosol characteristics and particle production in the upper troposphere over the Amazon Basin. *Atmos Chem Phys* 2018; **18**: 921–961.